\documentclass{elsart}  
\usepackage{epsfig,rotating,amstex,amssymb}

\def\arctanh{{\,\rm arctanh}\,}
\def\arccoth{{\,\rm arccoth}\,}
\newcommand{\lsim}   {\mathrel{\mathop{\kern 0pt \rlap
  {\raise.2ex\hbox{$<$}}}
  \lower.9ex\hbox{\kern-.190em $\sim$}}}
\newcommand{\gsim}   {\mathrel{\mathop{\kern 0pt \rlap
  {\raise.2ex\hbox{$>$}}}
  \lower.9ex\hbox{\kern-.190em $\sim$}}}

\begin{document}
\begin{frontmatter}

\title{Radiative neutrino decays in very strong magnetic fields}

\author{M. Kachelrie{\ss} and G. Wunner}
\address{Theoretische Physik I, Ruhr-Universit\"at Bochum,
         D--44780 Bochum, Germany}
         
\thanks{To appear in Physics Letters B.}

\begin{abstract}
The radiative decay $\nu_H \to \nu_L + \gamma$
of massive neutrinos is analyzed in the framework of the standard
model with lepton mixing for very strong magnetic fields 
$B\gg B_{\rm cr}=m_e^2/e\sim 4.14\times 10^{13}\:$G.
The analysis is based on the approximate decay amplitude obtained
by Gvozdev {\it et al.\/} Numerical results as well as analytical
approximations for the decay rate are obtained
for energies of the initial neutrino below and {\em above} the 
electron-positron pair creation threshold $2m_e$. 
\end{abstract}

 
\end{frontmatter}

\section{Introduction}

In the Standard Model of electroweak interactions (SM) the neutrino masses
are set to zero ``by hand''. While all terrestrial experiments are
consistent with this assumption, it is now clear that the solution of
the solar neutrino problem requires nonzero neutrino masses and neutrino
mixing \cite{be96d}. Another motivation for neutrino masses comes from
cosmology with the need of a hot dark matter particle in the few eV mass
range or of a heavy ($\sim\:$MeV) unstable particle 
\cite{taup}. In both cases, the $\tau$-neutrino is an ideal candidate. 
However, astrophysics and cosmology provide also stringent limits for
neutrino masses and lifetimes \cite{books}. In particular, the requirement
that the neutrino energy density does not overclose the Universe
restricts the sum of the masses of all stable neutrino species to be less than 
$92\, \Omega_\nu h^2 \:$eV, where $\Omega_\nu h^2 \lsim 1$.

In extended SMs with neutrino mixing, neutrinos
can decay by the process $\nu_H \to \nu_L + \gamma$ and, if the mass
$m_H$ of the heavy neutrino is higher than $2m_e$, additionally 
by $\nu_H \to \nu_L + e^- + e^+$. The resulting lifetimes
$\tau=\Gamma^{-1}$  are, 
although strongly model-dependent, extremely long. In the simplest
case where right-handed singlet fields $N_{lR}$ are added to the SM
the decay rate of the resulting Dirac neutrinos is \cite{books}
\begin{equation}  \label{tau_SM}
 \Gamma_{\rm SM} = 
                   \frac{\alpha}{2}\: \left( \frac{3G_F}{32\pi^2} \right)^2 
                   \left( \frac{m_H^2 - m_L^2}{m_H} \right)^3
                   \left( m_H^2 + m_L^2 \right)
                   \left| \sum_{l=e,\mu,\tau} U_{l H} U_{l L}^\ast 
                                   \:\frac{m_l^2}{m_W^2}
                   \right|^2 \:,  
\end{equation}
where $m_H,m_L$ are the masses of the heavy and of the light neutrino,
respectively.
If we assume that the $\tau$-lepton dominates the sum over the internal
leptons $l$ and that the elements of the mixing matrix $U$ are of order
unity, eq. (\ref{tau_SM}) yields a lifetime 
$\mathcal {O}(\tau_{\rm SM}) = 10^{29}$
years for $m_H=30\:$eV and $m_L \ll m_H$.

Since $\tau_{\rm SM} \gg T_{\rm Universe}$, neutrinos in the
SM with lepton mixing
can be treated as stable and the limit $m_\nu \lsim 92\:$eV applies.
The principal reason for the extremely small decay widths of
neutrinos is their small mass. Obviously, the decay rate $\Gamma$
for the radiative decay has to be proportional to $\alpha G_F^2$, and
since for $m_H \gg m_L$ the only energy scale available is the mass of
the heavy neutrino, it follows 
$\Gamma\propto \alpha G_F^2 m_H^5$ on purely dimensional grounds. 

This changes drastically if an external field is present. Then,
in the low-field limit $B\ll B_{{\rm cr},i}= m_i^2 /e$ 
the decay rates in magnetic fields $B$ depend 
only on the dynamical field parameters 
\begin{equation}
 \chi_i = m_i^{-3} \sqrt{ \left( p_\mu eF^{\mu\nu} \right)^2 } =
          \frac{p_\bot}{m_i}\:\frac{B}{B_{{\rm cr},i}} ,
\end{equation}
where $p_\bot$ is the momentum perpendicular to $B$ of the initial
particle and $B_{{\rm cr},i}$ are the critical fields of the charged
particles, while in the strong field limit 
$B \gg B_{{\rm cr},i}= m_i^2 /e$ the decay
rates can depend separately on $p_\perp$ and $B$ \cite{ni64/te94}. 
In both cases, the essential point
is that  the energy scale is set by the charged particles
and not by the mass of the decaying particle.

The best example for this mechanism is photon splitting $\gamma
\to 2 \gamma$. This process is forbidden in vacuum not only due
to the Furry theorem but also due to the zero mass of the photon. 
In an external field photon splitting is allowed and the energy scale
is set by the charged particles running in the loop.
For, e.g., $B\ll m^2_e /e$ and $\omega\ll 2m_e$ the dominating
contribution to the loop is that of the electron, and
the decay rate is given by \cite{ad71}
\begin{equation}
 \Gamma \propto \alpha^3 \:\frac{m_e^2}{\omega} \,
                             \left( \frac{k_\bot}{m_e} \right)^6 
                             \left( \frac{eB}{m_e} \right)^6 
        = \alpha^3 \:\frac{m_e^2}{\omega} \: \chi_e^6 \:.
\end{equation}

Consequently, the decay of all light particles should be enhanced in
strong external fields. For the example of the radiative decay of
neutrinos this was shown for various external fields
configurations by  Gvozdev {\it et al.\/} \cite{gv92,gv94/va95}. 
In the case of a
weak magnetic field $B\ll B_{\rm cr}$, they derived approximate decay
rates valid for arbitrary momentum $p_\perp$ of the initial neutrino.
However, in the strong field limit $B\gg B_{\rm cr}$, they gave 
an asymptotic expression for the decay width valid only for 
$p_\perp \ll 2m_e$ \cite{gv92}. 
 
Recently several authors proposed scenarios in which a strong
magnetic field with $B\gg m_e^2 /e$ in the early Universe was
created. Among the ideas considered are the production of a magnetic field
of $\mathcal{O}(B)= 10^{-8}M_{\rm GUT}^2 /e \approx 10^{43}\:$G 
by the ferromagnetic Yang-Mills vacuum \cite{en94}, 
of $\mathcal{O}(B)= m_W^2 / e \approx 10^{24}\:$G during the electroweak phase
transition \cite{va91}, 
of $\mathcal{O}(B)= (\lambda' M_{\rm Pl} m \phi_0)^{2/3}  \approx 10^{48}\:$G
during hybrid inflation, where $\phi$ is the inflaton 
field with mass $m$ and Higgs-coupling $\lambda'$ \cite{da95}, 
inflation in string cosmology \cite{ga95}, etc.  
Subsequently, several authors used
nucleosynthesis to derive bounds on primordial magnetic fields 
\cite{ns,ch96,gr96}.
In these works, the effect of the magnetic field 
on the weak
interaction rates $n \rightleftarrows p + e^- + \bar\nu_e$
and on the expansion rate of the Universe was taken into account 
but the neutrinos were treated as stable.
The limits obtained vary between $B \approx 2 \times 10^{13}\:$G in 
ref. \cite{ch96} and $B \gsim 1 \times 10^{15}\:$G in
ref. \cite{gr96} at the beginning of nucleosynthesis ($T\approx 1\:$MeV). 

Obviously, if the neutrino lifetime becomes of the order of the age 
of the Universe at the time of nucleosynthesis or before,
they can no  longer be treated as stable.
Since the effect of a magnetic field is equivalent to an increase of
the effective number of neutrino species $N_\nu$, while the decay of
a heavy neutrino before nucleosynthesis decreases $N_\nu$ from three
to two, the limits obtained in ref. \cite{ch96,gr96} could be even
weakened by taking into account neutrino decays.

%%%%%%%%%%%%%%%%%%%%%%%%%%%%%%%%%%%%%%%%%%%%%%%%%%%%%%%%%%%%%%%%
\section{Decay rate of $\nu_H \to \nu_L + \gamma$}

We consider the radiative decay of a heavy neutrino $\nu_H$ into a
lighter  neutrino $\nu_L$ and a photon $\gamma$ in a strong
magnetic field $B \gg B_{\rm cr}=m_l^2 /e$. Without loss of
generality, we can
choose the magnetic field as $\vec B = B \vec e_z$ and the
three-momentum  $\vec p$ of the heavy neutrino perpendicular to $\vec B$, 
e.g. $p^\mu = (E,p_{x},0,0)$. The momenta of the light neutrino
and of the photon are denoted by 
$p^{\prime\mu} = ( E^\prime, \vec p^\prime)$ and
$k^\mu = (\omega, \vec k)$; the components perpendicular to $B$ have
the index $\bot$. If the energy of the heavy neutrino is
small compared to the mass of the $W$-boson, $E \ll m_W$, the
four-fermion interaction can be used and the decay is described by the
Feynman diagram shown in Fig.~1. In this limit,
Gvozdev {\it et al.\/} \cite{gv92} found for the  matrix element 
$\mathcal{M}$ of the process $\nu_H \to \nu_L + \gamma$  
\begin{equation}   
 \mathcal{M} ( B \gg B_{\rm cr}) = 
  - \frac{e}{24\pi^2} \: \frac{G_F}{\sqrt{2}} \: C \:
    \sum_{l=e,\mu,\tau} U_{Hl}U_{Ll}^\ast \: \frac{eB}{m_l^2} \: f(x) \,, 
%    \langle \frac{eB}{m_l^2} f \left(x \right) \rangle \; ,
\end{equation}
where an effective matrix element 
\begin{equation}   
 C  =  k_\bot^2 
         \left( \varepsilon^\ast_\mu \tilde\Phi^{\mu\nu} j_\nu
         \right)
         -    
         \left( k_\mu \tilde\Phi^{\mu\nu} j_\nu \right)
         \left( k_\mu \tilde\Phi^{\mu\nu} \varepsilon^\ast_\nu
         \right) 
\end{equation}
and a function $f(x)$ depending only on the ratio $x= 4m_l^2 / k_\bot^2$
were introduced. %The bracket $\langle\ldots\rangle$ denotes a 
%summation over the different internal leptons, 
%\begin{equation}
% \langle \frac{eB}{m_l^2} \: f(x) \rangle =
% \sum_{l=e,\mu,\tau} U_{Hl}U_{Ll}^\ast \: \frac{eB}{m_l^2} \: f(x) \,, 
%\end{equation}
%
The other abbreviations have the following meaning:
The neutrino current $j$ is given by 
$j_\mu = \bar\nu_L (p_2)\gamma_\mu (1+\gamma^5)\nu_H (p_1)$,
$\tilde\Phi_{\mu\nu}= \tilde F_{\mu\rho}\tilde F^\rho_{\;\nu} / B^2
- \tilde F_{\mu\nu} / B$ and
$\tilde F$ is the dual field tensor.
For $x>1$ the function $f$ is given by
\begin{equation}   \label{f>1}
 f(x) = \frac{3}{2} x \left( x \Omega \arctan\Omega -  1 \right)
\end{equation}
with
\begin{equation}
 \Omega  = \frac{k_\bot}{\sqrt{4m^2 - k_\bot^2}} , 
\end{equation}
where we corrected a typographical error (in the argument
of $\arctan$) of ref. \cite{gv92}; a check of the relative signs in
eq. (\ref{f>1}) is given by the requirement $\mathcal{M}\to 0$ for
$\omega\to 0$. 
The function $f$ can be analytically continued in the 
complex $k_\bot^2$ plane. Then, for $x<1$, the real part
of $f$ follows simply as
\begin{equation}   
 \Re (f) =  \frac{3}{2} x \left( - x \Phi \arctanh\Phi -  1 \right)
\end{equation}
with
\begin{equation}
 \Phi = \frac{k_\bot}{\sqrt{k_\bot^2 - 4m^2}}    \,, 
\end{equation}
while the imaginary part of $f$ is given by the dicontinuity across
the branching cut starting at $k_\bot^2 = 4m_e^2$, viz. 
$2i \Im (f)= f(k_\bot^2 +i\varepsilon )-f(k_\bot^2 -i\varepsilon )$,
\begin{equation}   
 \Im (f) = \frac{3}{4}\: i\pi x^2 \Phi   \,.
\end{equation}
Using 
$\arccoth z = \half \ln \frac{1+z}{1-z} + \half\pi i$ for $\Im (z) <0$,
one sees that $f(x)$ for $x<1$ given in ref. \cite{gv92}
coincides---except for the sign of the term $3/2 x$---with the real
part given by us\footnote{This sign error has also been corrected by
  Gvozdev {\it et al.\/} in ref. \cite{va96}.}, 
but that the authors of ref. \cite{gv92} missed to
evaluate the imaginary part of $f$.

In the following, we restrict ourselves to the case that the internal
loop is dominated by the electron, i.e. that $m_e^2 /e \ll B \ll
m_\mu^2 /e$. In this case, the contribution of  additional
charged particles, e.g. from an enlarged Higgs sector, to the SM can be
neglected. Therefore the decay rates will hold in the limit $m_H \gg
m_L$ not only for the SM with lepton mixing but also for more general
models of massive neutrinos\footnote{If neutrinos are Majorana
  particles, the decay rates should be divided by two.}. Moreover, we
set $U_{Hl}U_{Ll}^\ast =1$ for simplicity. 

The neutrino decay rate $\Gamma$ is given by 
\begin{equation}  \label{rate}
 \Gamma = \frac{1}{16\pi E^2}\int_0^{E} {\rm d}\omega \, 
          |\overline{\mathcal{M}}|^2.
\end{equation}
Averaging $|C|^2$ over spins, we obtain
\begin{eqnarray}
 |\overline{C}|^2 & = & 16 \big[ \: k_\bot^2  
       \big( 2 E \omega^2 E^\prime - 2 E \omega k_z E^\prime 
             +  E k_z^2 p_{z}^\prime +  E p_{z}^\prime \omega^2 \big) 
\nonumber\\ & & 
       + 2 \big( E^\prime \omega - E^\prime k_z + p_{z}^\prime \omega 
                 + p_{z}^\prime k_z \big)
           \big( k_z - \omega \big) E \omega k_z 
       + E p_{z}^\prime k_\bot^2
\nonumber\\ & & 
       +  \big( E E^\prime - p_{x} p_{x}^\prime \big)
\nonumber\\  & & 
    \big( -  k_x^{2} \omega^{2} - k_x^{2} k_z^{2} -  k_y^{2} \omega^{2} 
         -  k_y^{2} k_z^{2} + E \omega^{3} k_z E^\prime 
\nonumber\\  & &  
         - 2 E \omega^{2} k_z^{2} E^\prime  +  E \omega k_z^{3} E^\prime 
         +  E \omega^{3} k_z p_{2}^\prime - E \omega k_z^{3} p_{z}^\prime 
    \big) \big]  \: .
\end{eqnarray} 
In the limit $E \gg m_H, m_L$, we can set $E \approx p$ and 
$E^\prime \approx p^\prime$. 
Therefore the three particles propagate collinearly
and the above expression reduces to 
\begin{equation}  \label{C-app}
 |\overline{C}|^2 =  32 \omega^4 E E^\prime \:. 
\end{equation}

The differential decay rate $\d\Gamma / \d\omega$ 
can be computed only for $k_\bot^2 \neq 2m_e$. 
Moreover, as the singularities of $|f(x)|^2$ at $k_\bot =2m_e$
are not integrable, the total decay rate $\Gamma$ above the pair
creation threshold is ill-defined. Physically speaking, we have not 
taken into account the finite lifetime of neutrinos and photons in
magnetic fields. Usually, the finite lifetime is incorporated
by the replacement of the energies $E$ by complexified energies 
$E -\frac{1}{2}i\Gamma(E)$ in the denominators, where $\Gamma(E)$ is
the decay width of the state with energy $E$.  
We note that, although this replacement seems natural, it is not unambiguous.
To proceed, we assume that the decay widths of the neutrinos are
neglectable compared to those of photons
and insert the total decay width 
\begin{equation}
 \Gamma^{\rm tot}= \Gamma_{\rm pair} (\omega)+ \Gamma_{\rm splitt} (\omega)
\end{equation}
in the energy denominators of $\Omega$ and $\Phi$. 
Here, $\Gamma_{\rm pair}$ denotes the decay width of the process 
$\gamma \to 2e$ \cite{da83} and $\Gamma_{\rm splitt}$
the decay width of the process 
$\gamma \to 2\gamma$ \cite{ba96/ka96b}.

Let us now consider the two analytically tractable limits 
$p_{\bot} \ll 2m_e$ and $p_{\bot} \gg 2m_e$. 
The function $f(x)$ behaves as 
$f(x) \sim \frac{3}{2}\:x$ for $x\to 0$ and goes to
$1$ for $x\to\infty$. Inserting this and eq. (\ref{C-app}) into
eq. (\ref{rate}), we obtain
\begin{eqnarray}      \label{E<<m} 
 \Gamma & =  & \frac{\alpha G_F^2}{15\cdot 288 \pi^4 \, E} 
               \left( E \sin\theta \right)^6 
               \left( \frac{B}{B_{\rm cr}} \right)^2 
\quad {\rm for} \quad m_H \ll p_{\bot} \ll 2m_e
\\                     \label{E>>m} 
 \Gamma & = & \frac{\alpha G_F^2 m_e^5 }{8 \pi^4 \, E} \: \chi_e^2
%E_1  \left( \frac{B}{B_{\rm cr}} \right)^2 
\quad {\rm for} \quad 2m_e \ll p_{\bot} \ll m_W  \,,
\end{eqnarray}
where $\theta$ denotes the angle between the magnetic field $\vec B$
and $\vec p$.
The first approximation was already derived in Ref. \cite{gv92}.
%The second one -- with a slightly different coefficient --
%can also be obtained using the semi-classical method of Ref.
%\cite{semicl}. 
Note that the energy dependence of the total decay
rate is the same as that for photon splitting.

We now discuss our numerical results, which were all computed using
the approximation eq. (\ref{C-app}) and with $B=10\,B_{\rm cr}$,
$\theta =90^\circ$.
In Fig. 2 the differential decay rate $\d\Gamma/\d\omega$
(normalized to unity when integrated over the abscissa)
is shown for energies of the initial neutrino below (left) and above
(right) the pair-creation threshold as a  
function of the energy $\omega / E_1$ of the emitted photon.
The energy distribution between the photon and the light neutrino
becomes more and more asymmetric for $E \to 2m_e$, and finally, for 
$E > 2m_e$, the resonance for $\omega =2m_e$ appears. For $\omega =
100 m_e$, the differential decay rate has a plateau similar to the
case of photon splitting.

A comparison between the total decay rate eq. (\ref{rate}),
its low-energy approximation eq. (\ref{E<<m})  and its high-energy 
approximation eq. (\ref{E>>m}) is made in Fig. 3.  
It can be seen that the exact rate is well reproduced
by the approximations except for small region around $2m_e$.

%In Fig. 4 the lifetime $\tau$ of the heavy neutrino is shown as a
%function of
Finally, we have computed the lifetime $\tau$ of the heavy neutrino as
a function of the age $t$ of the Universe. Hereby we have assumed that
the magnetic field
scales like $B(t)= B(t_0) (t_0 /t)^2$, and used as the starting value
$B_0$ the limits obtained in ref. \cite{ch96} and \cite{gr96}
for the onset of nucleosynthesis at $t_0 \approx 1$s. If the lifetime
is below %the line 
$t=\tau$ and the temperature $T$ of the
Universe lower than $m_H$, the heavy neutrino
decouples. Decoupling occurs for $m_H > 11\:$MeV using  $B_0 =2 \times
10^{13}\:$G and for $4\:$MeV using $B_0 = 1 \times 10^{15}\:$G. 
Therefore, the frequently discussed 
$\tau$-neutrino with $m_H \sim\:$MeV \cite{MeV}
is possible in this scenario
without introduction of new interactions.

%%%%%%%%%%%%%%%%%%%%%%%%%%%%%%%%%%%%%%%%%%%%%%%%%%%%%%%%%%%%%%%%%%%%%%%
\section{Summary}

We have presented the first analysis of the rate
of radiative neutrino decay {\em not\/} restricted to momentum of the initial 
neutrino $p_\bot \ll 2m_e$ in the limit of strong magnetic fields
$B\gg B_{\rm cr}$. We have derived two simple approximations
for the total decay rate valid for $p_{\bot} \ll 2m_e$ and
$p_{\bot} \gg 2m_e$
in this limit. The  decay rates obtained are extremely enhanced by the
strong magnetic field. In particular, if a strong primordial magnetic
field existed in the early Universe, it is possible that a heavy
neutrino with only standard model interactions may have decayed before
the onset of nucleosynthesis.

%%%%%%%%%%%%%%%%%%%%%%%%%%%%%%%%%%%%%%%%%%%%%%%%%%%%%%%%%%%%%%%%%%%%%%% 
\begin{ack}
We wish to thank Prof. V. Berezinsky for stimulating discussions
which initiated this work and  Dr. Vassilevskaya for sending us
ref. \cite{va96} prior to publication.
\end{ack}

\end{document}